\documentclass[aps,pra,epsfigure,twocolumn]{revtex4}
\usepackage{dcolumn}
\usepackage{bm}
\usepackage{graphicx}
\usepackage{amsmath}
\usepackage{latexsym}
\usepackage{amsfonts}
\usepackage{amssymb}
\usepackage{array}
\usepackage{epsfig}
\usepackage{times}

\newcommand{\ket}[1]{\left\vert#1\right\rangle}

\newcommand{\bra}[1]{\left\langle#1\right\vert}

\newcommand{\nbar}{\overline{n}}

\begin{document}

\title{Violations of Bell's inequality for Gaussian states with
homodyne detection and nonlinear interactions}

\author{M. Paternostro$^1$, H. Jeong$^{2,3}$, and T. C. Ralph$^3$}

\affiliation{$^1$School of Mathematics and Physics,
Queen's University, Belfast BT7 1NN, United Kingdom \\
$^2$Center for Subwavelength Optics and Department of Physics and Astronomy, Seoul National
University, Seoul, 151-742, South Korea\\
$^3$Center for Quantum Computer Technology, Department of Physics,
University of Queensland, St Lucia, Qld 4072, Australia
}

\begin{abstract}
We show that homodyne measurements can be used to
demonstrate violations of Bell's inequality with Gaussian states, when the
local rotations used for these types of tests
are implemented using nonlinear unitary
operations. We reveal that the local
structure of the Gaussian state under scrutiny is crucial in the
performance
of the test. The effects of finite detection efficiency is thoroughly
studied and shown to only mildly affect the revelation of Bell violations.
We speculate that our approach may be extended to other applications such as entanglement
distillation
where local operations are necessary elements besides quantum
entanglement.

\end{abstract}

\date{\today}

\maketitle

Violation of Bell's inequality~\cite{bell-original,bell},
which means failure of local realism,
is perhaps the most profound
yet controversial feature of quantum mechanics.
It was Einstein, Podolsky and Rosen (EPR)'s work which
challenged the completeness of quantum mechanics
as a theory~\cite{epr}.
Although the
original formulation of EPR's paradox involved the state
of a bipartite system having a continuous spectrum, Bohm's
version of the problem~\cite{bohm} and the seminal work by
Bell~\cite{bell-original,bell}
moved the debate towards its discrete version, which quickly
became the paradigm in the physics community.

The proved experimental handiness of continuous variable (CV) systems,
epitomized by proposals and realizations of schemes for quantum
teleportation
\cite{teleporto}, among other examples, has
redirected considerable interest towards the investigation of
Bell's inequality with these states.
It has been known since Bell that 
the original EPR state allows a realistic description in terms
of the canonically conjugated variables of
position and momentum because its Gaussian
Wigner function can be used as a classical
probability distribution for a hidden variable model \cite{bell}.
Later, however, it has been
proven that a two-mode squeezed vacuum state, whose Wigner function is a
Gaussian,
can violate Bell's inequality,
although non-optimal, in the joint measurement of phase-space
displaced parity operator~\cite{BW}.
The loophole here is that the inclusion of photon counting measurements 
(necessary for parity determination) negates  a reakistic interpretation of the Wigner function 
\cite{ralphetal}.
 This result has triggered extensive
investigation
on the interplay between measurements and Gaussian character of CV states
in Bell's inequality tests.
It has been found that non-Gaussian CV entangled states can be used to
demonstrate
violations of Bell's inequality
by means of Gaussian measurements ({\it i.e.}
measurements that preserve the Gaussian nature of a state,
such as homodyne detection)~\cite{peculiar}.
It has also been shown that a {\it non-deterministic} Bell's inequality test
can be devised using Gaussian CV states
to show Bell violations when non-Gaussian conditioning measurements are
combined with homodyne measurements \cite{NS}.
However, in virtue of Bell's argument
for a realistic description of an EPR state,
it is well known that Bell's inequality tests
in which both states and measurements are Gaussian, are
destined to satisfaction of Bell's inequality~\cite{bell}.
More practically, this ``no-go'' result seems paired 
with the impossibility of entanglement
distillation for Gaussian states under Gaussian-preserving
operations~\cite{fiurasek},
which in turn limits the
implementation of efficient quantum repeaters for
Gaussian states~\cite{qr}.

\begin{figure}[b]
\centerline{\includegraphics[width=7.8cm,height=5.0cm]{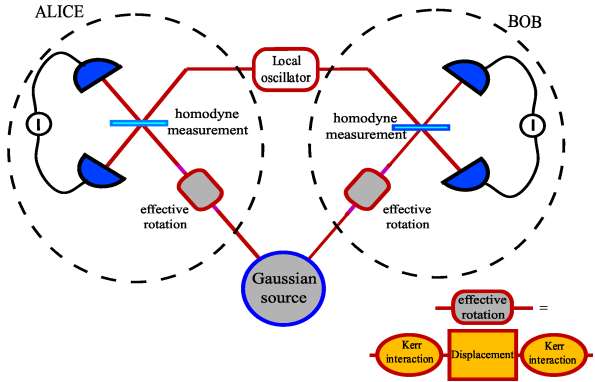}}
\caption{Schematic of a Bell-CHSH inequality test with Gaussian states
and homodyne measurements. A quantum correlated two-mode Gaussian states
is
produced at a source and distributed to Alice and Bob who perform local
effective rotations and homodyne measurements.
The effective rotations are physically implemented by cascading Kerr-type
nonlinearities and phase-space displacement operations, as shown in the
inset.
The Gaussian state produced
by the source can be either a two-mode squeezed state or the state
resulting from the superposition,
at a $50:50$ beam splitter, of vacuum and a single-mode squeezed state.}
\label{fig:scheme}
\end{figure}

In this paper, we describe a different angle on
the aforementioned problems. We examine
Clauser-Horne-Shimony-Holt (CHSH)'s version of Bell's
inequality~\cite{exp}
and show its violation by Gaussian states subjected to nonlinear local
unitary operations
and homodyne measurements. In order to efficiently illustrate our
findings,
which we describe with respect to routinely generated entangled Gaussian
states,
it is convenient to first address a
situation where the local operations
required for our Bell's inequality test are treated as ideal single qubit
rotations
\cite{stobinska}
for coherent-state qubits \cite{ideal,Ralph03}. We then
present physical transformations \cite{ideal}
which are able to reproduce the core features of
the idealized case (cfr. Fig.~\ref{fig:scheme}).
In both ideal and physical cases, a considerable
(although non-optimal) degree of violation of the Bell-CHSH test is
found,
resilient to imperfections at the detection stage.
{Our study shows that the local unitary operations, the importance
of which have not previously been carefully examined,
may play a key role in utilizing Gaussian states for
Bell's inequality tests and, possibly, other applications such as
entanglement distillation. Our protocol is shown to be 
quite resilient to spoiling effects such
as non-zero temperatures and low detection efficiency.
We also stress that our scheme makes use of Gaussian 
squeezed states, which are resources
routinely employed in all-optical experiments dealing 
with continuous variables. This distinguishes our proposal from
the case of entangled coherent states being used for Bell's inequality
tests (such as in Ref.~\cite{stobinska}), which are non-Gaussian and more demanding to produce.

The remainder of this paper is organized as follows.
In Section~\ref{test} we describe the formal approach to
our Bell's inequality test. We introduce the Gaussian states
we probe and the class of nonlinear unitary operations
we consider. This is achieved by first studying a sort
of idealized case and then moving to the real physical
situation. We show that Bell's inequality can be violated by Gaussian
states with homodyne measurements
using the specific class of operations.
In Section~\ref{robustness}
we study the effects of detection inefficiencies and show that they can
be counteracted by increasing the squeezing
of the initial resource. 
This very same strategy can be used in order to
cope with a mixedness initial state instead of a pure, ideal resource.
{This is consistent  with a previous result \cite{ourown}
using entangled thermal states \cite{jr06},
where the same effects could be achieved
by increasing the distance between the component thermal states \cite{ourown}.}
Section~\ref{testTMSS} shows that
the required level of squeezing to show Bell violations can be
considerably reduced
when another class of experimentally relevant Gaussian
states are used. Finally, in
Section~\ref{conclusions} we summarize our results and discuss their
physical implications.
}

\section{Test for Bell's inequality}
\label{test}

In Ref.~\cite{stobinska}, it was shown that a superposition of two
coherent states,
$|c_+\rangle=N_+(|\alpha\rangle+|-\alpha\rangle)$ with
the normalization factor $N_+$ and coherent states $|\pm\alpha\rangle$
of amplitudes $\pm \alpha$, when divided at a beam splitter,
violates Bell-CHSH inequality using homodyne measurements
and nonlinear interactions.
One can show that the fidelity $\cal F$
between a coherent-state superposition $|c_+\rangle$
and a single-mode squeezed state is very high when
$\alpha$ is relatively small (e.g. ${\cal F}\geq 0.99$ for $\alpha<0.75$).
This motivates us to first investigate violation of Bell's inequality
for single-mode squeezed states divided at a beam splitter. We shall later
study another set of Gaussian states which outperform the results for this
case.

Let us suppose that two parties, Alice and Bob,
share an entangled state generated using a single-mode squeezed vacuum
and a $50:50$ beam splitter~\cite{kokralph}. Analytically, the state can
be described by the
following Gaussian-weighted continuous superposition of coherent
states~\cite{papers}
\begin{equation}
\label{initial}
\ket{\xi}_{AB}={\cal N}\int{d}^2\alpha~{\cal G}(r,\alpha)
|\frac{\alpha}{\sqrt{2}},\frac{\alpha}{\sqrt{2}}\rangle_{AB},
\end{equation}
where ${\cal G}(r,\alpha)=\exp[{-(1-\tanh{r})\alpha^2/({2\tanh{r}})}]$,
$r$ is the squeezing parameter, $\alpha\in\mathbb{R}$ and ${\cal
N}=1/\sqrt{2\pi\sinh{r}}$
is the normalization factor. The class of nonlinear transformations we
consider can be
understood as an approximation of
the following rotations performed in the bidimensional space
spanned by the generic coherent state $\{\ket{\pm\beta}\}$
($\beta\in\mathbb{C}$)~\cite{stobinska}
\begin{equation}
\label{ideale}
\begin{split}
&\hat{R}_{j}(\theta)\ket{\beta}_j\rightarrow
\sin(2\theta_j)\ket{\beta}_j+\cos(2\theta_j)
\ket{-\beta}_j,\\
&\hat{R}_{j}(\theta)\ket{-\beta}_j\rightarrow
\cos(2\theta_j)\ket{\beta}_j-\sin(2\theta_j)\ket{-\beta}_j,
\end{split}
\end{equation}
where $\theta_j$ is the effective ``angle'' of such idealized rotations
and $j=A,B$ labels Alice's or Bob's site.
It should be noted that the ``idealized" transformation described in
Eq.~(\ref{ideale})
is {\it not} unitary (approximately unitary when $\beta$ is large)
so that it cannot be performed deterministically.
The actual physical local transformation using nonlinear interactions
will be considered later in this Section.

\begin{figure}[t]
\centerline{\scalebox{0.65}{\includegraphics{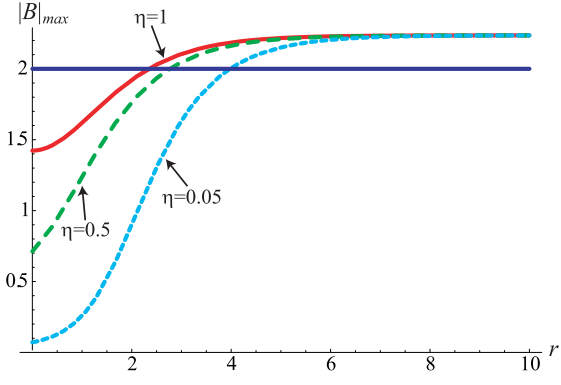}}
}
\caption{Bell test for ideal rotations.
The Bell-CHSH function is plotted against the squeezing parameter $r$
for three values of the detection efficiency $\eta$.
We show the case corresponding to ideal homodyne detection (solid line),
$\eta=0.5$ (dashed line) and $\eta=0.05$ (dotted line). The horizontal
line
shows the bound for local realistic theories.}
\label{fig:ide}
\end{figure}

After the application of the local operations~(\ref{ideale}) to their
respective mode, Alice and Bob perform bilocal homodyne measurements,
which result in the joint probability-amplitude function
\begin{equation}
C_{id}(\theta_A,\theta_B,x,y)\propto\langle{x,y}
|\hat{R}_A(\theta_A)\hat{R}_B(\theta_B)\ket{\xi}_{AB}
\end{equation}
with $\ket{x}$ ($\ket{y}$) the in-phase quadrature eigenstate of Alice's
(Bob's) mode. A sketch of the thought-experiment is presented in
Fig.~\ref{fig:scheme}.
In order to test CHSH version of Bell's inequality, we need
to construct a set of bounded dichotomic observables, which we do by
assigning value $+1$ to a homodyne-measurement's outcome larger than 0,
and $-1$ otherwise~\cite{peculiar}. With this, a joint probability of
outcomes
can be calculated as
\begin{equation}
P_{kl}(\theta_A,\theta_B)=
\int^{k_s}_{k_i} dx \int^{l_s}_{l_i} dy~|C_{id}(\theta_A,\theta_B,x,y)|^2,
\end{equation}
where the subscripts $k,l=\pm$ correspond to Alice's and Bob's assigned
measurements
outcomes $\pm{1}$ and the integration limits are such
that $+_s=\infty,~+_i=-_s=0$ and $-_i=-\infty$. We can now calculate
the Bell-CHSH function, $B(\theta_A,\theta_B,\theta'_A,\theta'_B)={\cal
C}
(\theta_A,\theta_B)+{\cal C}(\theta'_A,\theta_B)
+{\cal C}(\theta_A,\theta'_B)-{\cal C}(\theta'_A,\theta'_B)$,
where we have introduced the correlation function
\begin{equation}
\label{corre}
{\cal C}(\theta_A,\theta_B)=\sum_{k,l=\pm}P_{kk}
(\theta_A,\theta_B)-\sum_{k\neq{l}=\pm}P_{kl}(\theta_A,\theta_B).
\end{equation}
According to local-realistic theories, the Bell-CHSH inequality
$|B(\theta_A,\theta_B,\theta'_A,\theta'_B)|\le{2}$ holds.
Quantitatively, we have found that
\begin{equation}
\label{corre1M}
{\cal C}_{id}(\theta_A,\theta_B,r)\!=\!\frac{2\text{arctan}(\sinh{r})
\cos(4\theta_A)\cos(4\theta_B)}{\pi(1+\mathop{\sum}\limits_{j\neq{k}}
\sin(4\theta_j)[\frac{\sin(4\theta_k)}{2}+\sinh{r}])}
\end{equation}
with $j,k=A,B$ and the subscript $id$ is used in order to remind of
the idealized version of local operations being used. The behavior
of the numerically optimized Bell-CHSH function corresponding to
Eq.~(\ref{corre1M})
is shown by the solid curve in Fig.~\ref{fig:ide}, which demonstrates
that a
local realistic description of $\ket{\xi}_{AB}$ is impossible as the
squeezing
parameter for the initial single-mode state surpasses $\sim{2.1}$. The
degree
of violation of the Bell-CHSH inequality then reaches a maximum of
$\sim2.23$,
robustly against $r$.

Now that we have gained a quantitative picture of the behavior of the Bell
function under the class of formal operations and Gaussian measurements
considered in our work, it is time to provide a physically effective
description
of each rotation $\hat{R}_{j}(\theta_j)$. Such physical implementation
stems
from the observation made in Ref.~\cite{stobinska} that
Eqs.~(\ref{ideale})
can be approximated by a combination of single-mode Kerr interaction
$\hat{U}_{\text{Kerr}}=e^{-i\hat{H}_{\text{Kerr}}t}$ with
$\hat{H}_{\text{Kerr}}=\hbar\Omega(\hat{a}^\dag\hat{a})^2$
($\Omega$ being the strength of the non-linear coupling)
and displacement of amplitude $\varphi\in\mathbb{C}$,
$\hat{D}(\varphi)=e^{\varphi\hat{a}^\dag-\varphi^*\hat{a}}$.
Here $\hat{a}$ ($\hat{a}^\dag$)
is the annihilation (creation) operator of a field mode.
A single-mode Kerr interaction may be implemented, for example,
by nonlinear crystals \cite{j04,v04} while the displacement can be easily
performed via
a beam splitter with high transmittivity and a local oscillator.
In detail, the evolution induced by the effective rotations
$\hat{V}_j(\theta_j)=\hat{U}_{\text{Kerr}}\hat{D}
(i\theta_j/d)\hat{U}_{\text{Kerr}}$
on input coherent states $\ket{\pm\beta}_j$ is given by the following
expressions
($\beta=\beta_r+i\beta_i$ and $d\in{\mathbb R}$ determines the amplitde of
 the displacement)~\cite{stobinska}
\begin{equation}
\label{rotazioni}
\begin{split}
\hat{V}_j(\theta_j)\ket{\beta}_j&=\frac{1}{2}
\left\{e^{i\frac{\theta_j}{d}\beta_r}
(|{\beta+\frac{i\theta_j}{d}}\rangle_j+i|
{-\beta-\frac{i\theta_j}{d}}\rangle_j)\right.\\
&\left.+ie^{-i\frac{\theta_j}{d}\beta_r}
(|{-\beta+\frac{i\theta_j}{d}}\rangle_j+
i|{\beta-\frac{i\theta_j}{d}}\rangle_j)\right\},\\
\hat{V}_j(\theta_j)\ket{-\beta}_j&=\frac{1}{2}
\left\{ie^{i\frac{\theta_j}{d}\beta_r}(|{\beta
+\frac{i\theta_j}{d}}\rangle_j
+i|{-\beta-\frac{i\theta_j}{d}}\rangle_j)\right.\\
&\left.+e^{-i\frac{\theta_j}{d}\beta_r}(|{-\beta
+\frac{i\theta_j}{d}}\rangle_j
+i|{\beta-\frac{i\theta_j}{d}}\rangle_j)\right\}.
\end{split}
\end{equation}
Note that $\hat{V}_j(\theta_j)$
is unitary while $\hat{R}_j(\theta_j)$ is not strictly a unitary
operation.
The physical operation $\hat{V}_j(\theta_j)$ is a good approximation of
the ideal operation $\hat{R}_j(\theta_j)$ when the amplitudes of coherent
states
on which the operation is acted are large.
As seen in Eq.~(\ref{initial}), our squeezed state can be expanded in
terms
of coherent states with a Gaussian weight factor as a function of the
coherent amplitude.
When the squeezing $r$ is large, contributions of
coherent states of small amplitudes will become arbitrarily small.
This implies that as the squeezing $r$ becomes large, the results of the
Bell-CHSH inequality violation
using the ideal rotation $\hat{R}_j(\theta_j)$ should be closer to the
results
using the physical rotation $\hat{V}_j(\theta_j)$.

We then adjust our notation and indicate with
${C}_{ef}(\theta_A,\theta_B,x,y)=|\langle{x,y}
|\hat{V}_A(\theta_A)
\hat{V}_B(\theta_B)|\xi\rangle_{AB}|^2$
the probability of measuring the values $x$ and $y$ of the quadrature
variables at the homodyne detectors. The subscript clearly states that
this is the function associated with the use of physical effective
rotations. Quantitatively,
${C}_{ef}$ is easily found using the projection of a coherent
state onto a position quadrature eigenstate $\ket{x}$, which is given by
$\langle{x}|\beta\rangle={\pi^{-1/4}}e^{\sqrt{2}
i\beta_ix-\frac{1}{2}(x-\sqrt{2}\beta_r)^2
-i\beta_r\beta_i}$~\cite{kokralph}. We eventually obtain
\begin{equation}
\label{corrphys}
\begin{split}
&|{C}_{ef}(\theta_A,\theta_B,x,y)|^2
=\frac{1}{\pi}e^{-r-e^{-r}\cosh{r}(x^2+y^2)}
\left(e^{xy{e}^{-2r}}\times\right.\\
&\sin[\frac{\sqrt{2}(y\,\theta_A+x\,\theta_B)}
{d}]\!+\!e^{xy}\cos[\frac{\sqrt{2}
(y\,\theta_A-x\,\theta_B)}{d}]\Big).
\end{split}
\end{equation}

\begin{figure}[b]
\centerline{\scalebox{0.8}
{\includegraphics{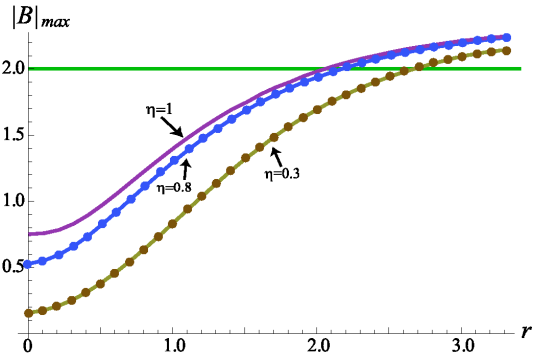}}}
\caption{The numerically optimized Bell function is plotted
against the squeezing parameter $r$ for the case of physical
effective rotations and three values of detection efficiency.
The horizontal line shows the bound for local realistic theories.
The actual value of $d$ is irrelevant, in this figure. The solid
line with $\eta=1$ embodies the ideal-detector case with
the other two cases of $\eta=0.8$ and $\eta=0.3$.}
\label{fig:physical2D}
\end{figure}

\begin{figure}[t]
\centerline{\scalebox{0.55}{\includegraphics{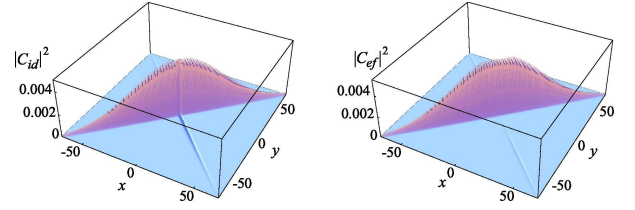}}}
\caption{We compare the behavior of the joint-probability functions
$|{C}_{id}(\theta_A,\theta_B,x,y)|^2$ and $|{\cal
C}_{ef}(\theta_A,\theta_B,x,y)|^2$
against the quadrature variables $x$ and
$y$ for $r=4$. The angles $\theta_{A,B}$
are those maximizing the corresponding
Bell-CHSH function. We thus have
$\theta_A\!=\!0.061$ and $\theta_{B}\!=\!0.182$
($\theta_A\!=\!-0.009$ and $\theta_{B}\!=\!0.004$)
for the leftmost (rightmost) plot. Moreover,
$\int\!\int{d}xdy|{C}_{ef}(\theta_A,\theta_B,x,y)|^2
\simeq\int\!\int{d}xdy|{C}_{id}(\theta_A,\theta_B,x,y)|^2$,
regardless of the domain of integration.
}
\label{fig:probs}
\end{figure}

We are now in a position to build up the Bell-CHSH function for
our Bell's inequality test in such physically effective case. Unfortunately,
producing an analytic result is rather demanding due to the semi-infinite
range of integrations over the quadrature variables $x$ and $y$, which
also
enter into the trigonometric functions in Eq.~(\ref{corrphys}),
required
in order to gather the joint probabilities $P_{kl}(\theta_A,\theta_B)$.
We have therefore performed the Bell's inequality test by numerically
evaluating the Bell-CHSH function for a set value of $d$ and by
scanning the squeezing parameter $r$. The results are shown by
the top-most curve in Fig.~\ref{fig:physical2D}, where violation
of local realistic theories starting from $r\gtrsim{2.1}$ is observed,
which is
in full agreement with the ideal-rotation case. Also, the degree of
violation is consistent between the two cases, $|B|_{max}$ being
$2.229$ at $r=3.3$.
Although the reproduction of the behaviour for large $r$ is
computationally demanding,
it is possible to perform a qualitative comparison between ideal
and effective case by looking at the corresponding joint probability
functions
$|C_{id}(\theta_A,\theta_B,x,y)|^2$ and Eq.~(\ref{corrphys}), evaluated
at the angles corresponding to the (numerically-optimized) associated
Bell-CHSH function.
This is done in Fig.~\ref{fig:probs}, where the clear similarity of the
two probability
functions ensures the closeness of the value of the corresponding
Bell-CHSH functions.

\section{Robustness to imperfections}
\label{robustness}

Although homodyne detectors
have rather high efficiencies, the violation of the Bell-CHSH inequality
by
$\ket{\xi}_{AB}$ is far from $2\sqrt 2$, the maximum given by Tsirelson's
bound~\cite{cirelson}. One might thus wonder whether even mild detection
inefficiencies are sufficient to wash out the Bell-CHSH inequality
violation unveiled in Fig.~\ref{fig:physical2D}.
An important issue to address is thus given by the effects of detection
inefficiencies.
As done before, we first gain an idea of the expected behavior by
studying the idealized picture.

In order to quantitatively assess this point, we have modeled the
imperfect homodyne detector onto which mode $j=A,B$ impinges as the
cascade
of a beam splitter of transmittivity $\eta$, mixing mode $j$ to an
ancillary
vacuum mode $a_j$, and a perfect homodyner. We are not interested in the
state
of the ancillae, which are discarded by tracing them out of the overall
state,
so that $|C_{id}(\theta_A,\theta_B,x,y)|^2$ is changed into
$\langle{x,y}|{\rm Tr}_{a_Aa_B}\psi(\theta_A,\theta_B)|x,y\rangle$
with
\begin{equation}
\begin{split}
&{\psi(\theta_A,\theta_B)}={\hat B}_{Aa_A}(\eta){\hat B}_{Ba_B}
(\eta)\hat{R}_A(\theta_A)\hat{R}_B(\theta_B)
\ket{\xi}_{AB}\!\bra{\xi}\\
&\otimes\ket{00}_{a_Aa_B}
\!\!\bra{00}\hat{R}^\dag_A(\theta_A)
\hat{R}^{\dag}_B(\theta_B){\hat B}^\dag_{Aa_A}
(\eta){\hat B}^\dag_{Ba_B}(\eta),
\end{split}
\end{equation}
where the beam splitter operation between
modes $j$ and the corresponding ancilla ${a}_j$
is defined as
${\hat B}_{ja_j}(\zeta)=\exp[{\frac{\zeta}{2}
({\hat a}_j^\dagger {\hat b}_{a_j}
-{\hat a}_j {\hat b}_{a_j}^\dagger)}]$ with
$\cos\zeta=\sqrt{\eta}$ and $\hat{b}_{a_j}$ being the annihilation
operator of
$a_j$~\cite{efficiency}. The remaining procedure for the construction of
the
appropriate Bell-CHSH function remains as described above. The final form
of
the correlation function, which now depends on the efficiency
as well, is obtained from Eq.~(\ref{corre1M}) by simply replacing
$\arctan(\sinh{r})\rightarrow\arctan
(\frac{\eta{e}^r\sinh{r}}{\sqrt{1+2\eta{e}^r\sinh{r}}})$. The behavior
of the associated Bell function is shown, for two values of $\eta$, in
Fig.~\ref{fig:ide}.
We observe a rather striking robustness of the Bell function with
respect to the
homodyners' inefficiency: Even severely inefficient homodyne detectors
would be able
to unveil Bell-CHSH inequality violations with a state which is
initially squeezed enough.
By simply increasing the squeezing of the input state, one can
compensate the effects
of detection inefficiencies. Although for small values of $\eta$, the
required squeezing
factor becomes prohibitively large, the trend revealed by the ideal
case leaves quite
a few hopes for the physical effective one as well.
In fact, such robustness persists when the local
operations~(\ref{rotazioni}) are used,
as shown in Fig.~\ref{fig:physical2D} for
$\eta=0.8$ and $0.3$ (chosen for easiness of representation).
The squeezing threshold at which the
Bell's inequality test starts to be violated increases only quite
slowly as the quality of the homodyne detectors is degraded.
In passing, we should stress that the beam-splitter model used for the
description
of an inefficient homodyne detector can be used in order to describe the
influences
of external zero-temperature reservoirs coupled to the correlated two-mode
state we
are studying. Thus, similar conclusions regarding the resilience of the
Bell-CHSH function to losses induced by a low-temperature environment can
be drawn.

\begin{figure}[t]
\centerline{\scalebox{0.7}{\includegraphics{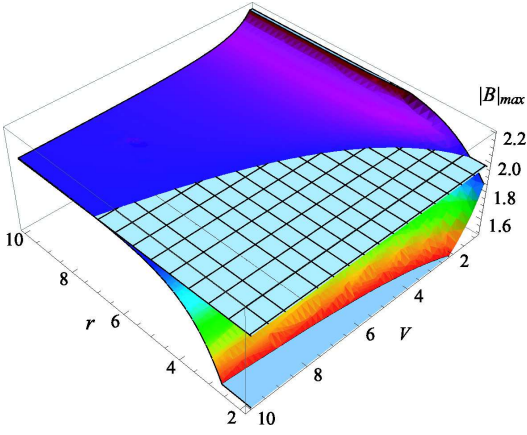}}}
\caption{Bell-CHSH test for an input squeezed thermal state
superimposed to vacuum at a $50:50$ beam splitter.
The Bell-CHSH function is plotted against $r$ and
$V=2\nbar+1$, {\it i.e.} the thermal variance of the state.
The horizontal plane shows the bound for local realistic theories.}
\label{fig:misto}
\end{figure}

We complete our study about the effects of imperfections by investigating
the case in which we
start with a mixed resource. This is practically quite relevant, given the
fact that, experimentally,
a single-mode squeezed thermal state is in general
produced instead of a pure single-mode squeezed vacuum state.
This is formally accounted for by considering the resource state
\begin{equation}
\label{squeezedthermal}
\rho^{st}_{AB}\!=\!\int{d}^2\alpha{\cal T}(\nbar,\alpha)\hat{S}_A(r)
|{\frac{\alpha}{\sqrt 2},\frac{\alpha}{\sqrt 2}}\rangle_{AB}
\langle{\frac{\alpha}{\sqrt 2},\frac{\alpha}{\sqrt 2}}|\hat{S}^\dag_A(r),
\end{equation}
where ${\cal T}(\nbar,\alpha)=
{e}^{-|\alpha-d|^2/\nbar}/\pi\nbar$ $(\alpha=\alpha_r+i\alpha_i)$
is the Glauber-Sudarshan function of a single-mode state at
thermal equilibrium with mean photon number $\nbar$ and displaced,
in phase space, by $d\in\mathbb{R}$~\cite{kokralph} while
$\hat{S}_A(r)=e^{\frac{r}{2}(\hat{a}^{\dag{2}}-\hat{a}^2)}$
is mode-$A$ squeezing operator. Eq.~(\ref{squeezedthermal})
results from superimposing at a $50:50$ beam splitter a squeezed
displaced thermal state of mode $A$ and the vacuum state of mode $B$.
It is straightforward to find that Eq.~(\ref{squeezedthermal}) can
be written as $\rho^{st}_{AB}=\int{d}^2\alpha\tilde{\cal T}(r,V,\alpha)
|{{\alpha}/{\sqrt 2},{\alpha}/{\sqrt 2}}\rangle_{AB}\!\langle{{\alpha}/
{\sqrt 2},{\alpha}/{\sqrt 2}}|$ with $V=2\nbar+1$ and
\begin{equation}
\tilde{\cal T}(r,V,\alpha)=\frac{2e^{-\frac{2\alpha^2_i}{e^{2r}V-1}
-\frac{2(\alpha_r-d)^2}{e^{-2r}V-1}}}{\pi\sqrt{V^2+1-2V\cosh(2r)}}.
\end{equation}
This state is then locally rotated and projected onto quadrature
eigenstates by means of homodyne measurements. Once more,
for clarity of our arguments, we refer to the case of ideal rotations.
The use of our formal procedure applied so far lead to the correlation
function
\begin{equation}
{\cal C}_{st}(\theta_A,\theta_B,r)\!=\!\frac{2\text{arctan}
(\frac{e^{r}-Ve^{-r}}{2\sqrt{V}})\cos(4\theta_A)\cos(4\theta_B)}
{\pi\left[1+\frac{\sin(4\theta_A)\sin(4\theta_B)}{V}+
\frac{2(\sin(4\theta_A)+\sin(4\theta_B)}{\sqrt{V^2+1+2V\cosh(2r)}}\right]}.
\end{equation}
Clearly, ${\cal C}_{st}\equiv{\cal C}_{id}$ when $V=1$, {\it i.e.}
when a pure state is generated. With this expression, one can easily build
up the Bell-CHSH function and test its behavior against the thermal
parameter $V$ and, as usual, the squeezing. The results are shown in
Fig. \ref{fig:misto}, where it is shown that it is enough to consider
a slightly more squeezed initial resource in order to counteract any
thermal effect. The same conclusions are reached by using the set of
nonlinear unitary transformations $\hat{V}_j(\theta_j)$,
although the analysis is largely numerical and more involved.

\section{
Improvement using two-mode squeezed states}
\label{testTMSS}

The required level of squeezing, {\it e.g.} $r\gtrsim 2$
for $\eta \geq 0.8$,
revealed in Figs.~2 and 3 to demonstrate Bell-CHSH inequality violations
is experimentally difficult to achieve using current technology.
In this Section, we show that this requirement can be
radically reduced by using another class of Gaussian states.
So far, we have investigated
the Bell's inequality test under nonlinear operations using the paradigmatic
source given by state
$\ket{\xi}_{AB}$. However, the behaviour of a Bell-CHSH function strongly
depends on intrinsic properties of the tested quantum correlated state.
In fact, this can be seen as the ``dual" of the well-known fact that the
same
bipartite entangled state behaves differently, in terms of Bell
inequality tests,
under different sets of local operations. Here, we are interested in
finding
out whether another realistic Gaussian resource is conceivable for the
violation
of our Bell-CHSH inequality when smaller values of $r$ are taken.
Our starting point is the observation~\cite{KSBK,bowen}
\begin{equation}
\hat{B}_{AB}(\frac{\pi}{2})\hat{S}_{A}(r)\ket{0,0}_{AB}
=\hat{S}_{A}(\frac{r}{2})\hat{S}_B(\frac{r}{2})
\hat{S}_{AB}(\frac{r}{2})\ket{00}_{AB},
\end{equation}
where we have used the single-mode squeezing operator
$\hat{S}_{j}(r)=\exp[\frac{r}{2}
(\hat{a}^2_j-\hat{a}^{\dag2}_j)]~(j=A,B)$ and its
two-mode version $\hat{S}_{AB}(r)=\exp[r(\hat{a}^\dag_A
\hat{a}^\dag_B-\hat{a}_A\hat{a}_B)]$.
Therefore, our resource $\ket{\xi}_{AB}$ is formally equivalent to a
two-mode
squeezed state that is also subjected to additional local squeezing
operation.
These latter are unable to change the nonlocal content of the state
being used
and could well be regarded as a pre-stage of the local actions
(comprising nonlinear
rotations and homodyne measurements) performed at Alice's and Bob's site
respectively.
We now remove them from the overall setup for Bell's inequality tests by
considering, instead
of Eq.~(\ref{initial}), the standard two-mode squeezed
vacuum~\cite{originalsqueezing}
\begin{equation}
\ket{\xi'}={\cal M}\int{d}^2\beta~{\cal G}'(r,\beta)\ket{\beta,\beta^*}
\end{equation}
with weight function~\cite{jlk}
\begin{equation}
{\cal G}'(r,\beta)=\exp[{-\frac{1-\tanh{r}
|\beta|^2}{{\tanh{r}}}}]
\end{equation} and normalization factor
${\cal M}=(\pi\,\sinh{r})^{-1}$. The adaptation of the formal
procedure described in our work to the use of this Gaussian resource
is quite straightforward. For the simple case of ideal local rotations,
the correlation function for joint outcomes at Alice's and Bob's site is
identical to Eq.~(\ref{corre1M}) with the replacement
$r\rightarrow{2r}$.
The violation of the local realistic bound occurs now for $r\sim{1}$
and
the entire Bell-CHSH function shown in Figs.~\ref{fig:ide} {\bf (a)} is
``shifted back"
on the $r$ axis accordingly
This effect
is the same when Eqs.~(\ref{rotazioni})
are used, although the form of $C_{ef}$ is too cumbersome to be shown
here.
For clarity, we note that a two-mode squeezed state of squeezing $r$
can be generated using two single-mode
squeezed states of the same degree of squeezing
and a beam splitter as
\begin{equation}
\hat{S}_{AB}(r)\ket{00}_{AB}=
\hat{B}_{AB}(\frac{\pi}{2})\hat{S}_{A}({r})
\hat{S}_B(-{r})\ket{0,0}_{AB}.
\end{equation}
This means that single mode squeezed states of $|r|\gtrsim 1$ ($\gtrsim
8.7$dB)
can be used as resources to show violations of Bell's inequality.
This makes our proposal closer to an experimental implementation
as such high levels of squeezing can be generated
(for example, up to 10dB \cite{squeezing})
using current technology.
On the other hand, the local nonlinear operations may be more demanding
and
various types of unitary interactions need to be investigated to improve
experimental feasibility of our approach.

\section{Conclusions}
\label{conclusions}

We have shown a way to unveil violations of Bell's inequality
for two-mode Gaussian states by means of
nonlinear local operations and
Gaussian homodyne measurements.
Besides its theoretical interest, which stays at the center of current
investigations on entangled CV systems and their fundamental features,
our study emerges as an appealing alternative to the current strategy
for Bell's inequality tests based on the use of appropriately de-Gaussified
resources and high-efficiency homodyning. Our proposal has been shown to be
robust against the inefficiency of the homodyne detection and mixedness
in the initial resource.
{This robustness is consistent with a previous study
using entangled thermal states \cite{ourown}.}

While the squeezing degree of $r \gtrsim 1$ required for resource
Gaussian states is possible to achieve
using present day technology, the strong nonlinear interactions
required to implement the local operations
may be more difficult to realize.
On the other hand, it is worth noting that there has been remarkable
progress to obtain strong nonlinear effects~\cite{v04,strong}.

There remains some interesting future work.
As the local operations used in our paper are not necessarily optimized
ones,
the research for more efficient local operations is desirable.
Since we have used nonlinear operations
to reveal violations of Bell's inequality
for Gaussian states and Gaussian measurements,
it is also natural to extend this investigation to entanglement
distillation protocols
for the Gaussian states.
Interesting open questions are therefore
whether there exist such entanglement distillation protocols
using the type of local operations
employed in this paper and how much they would be feasible and useful.

\acknowledgments

MP thanks M. S. Kim for useful discussions
and acknowledges the UK EPSRC for financial support (Grant number EP/G004579/1). 
This work was supported by
the Australian Research Council, Defence Science and Technology
Organization, and the Korea Science and Engineering Foundation 
(KOSEF) grant funded by the Korea government(MEST) (R11-2008-095-01000-0).


\begin{thebibliography} {99}
\bibitem{efficiency} For simplicity of presentation,
the detection efficiencies are taken to be equal for both the homodyners,
an assumption
that can be easily relaxed.
\end{thebibliography}
\end{document}